\documentclass[12pt,preprint]{aastex}

\slugcomment{}

\shorttitle{Galactic spectral index} \shortauthors{Tartari et al.}

\begin{document}


\title{TRIS III:  \\
    the diffuse galactic radio emission at $\delta=+42^{\circ}$}


\author{A. Tartari, M. Zannoni\altaffilmark{1}, M. Gervasi\altaffilmark{1,2} , G. Boella\altaffilmark{2}, and G. Sironi\altaffilmark{1}}
\affil{Physics Department, University of Milano Bicocca, Piazza della Scienza 3, I20126 Milano Italy}

\email{mario.zannoni@mib.infn.it}

\altaffiltext{1}{also Italian National Institute for Astrophysics, INAF, Milano.}

\altaffiltext{2}{also Italian National Institute for Nuclear Physics, INFN, Milano-Bicocca.}


\begin{abstract}

We present values of temperature and spectral index of the galactic diffuse radiation measured at 600 and 820 MHz along a 24 hours right ascension circle at declination $\delta = +42^{\circ}$. They have been obtained from a subset of
absolute measurements of the sky temperature made with TRIS, an experiment devoted to the measurement of the Cosmic Microwave Background temperature at decimetric-wavelengths with an angular resolution of about $20^{\circ}$.

Our analysis confirms the preexisting picture of the galactic diffuse emission at decimetric wavelength and improves the
accuracy of the measurable quantities. In particular, the signal coming from the halo has a
spectral index in the range $2.9-3.1$ above 600 MHz, depending on the sky position. In the disk, at TRIS angular resolution, the free-free emission accounts for the $11\%$ of the overall signal at 600 MHz and $21\%$ at 1420 MHz.
The polarized component of the galactic emission, evaluated from the survey by Brouw and Spoelstra,
affects the observations at TRIS angular resolution by less than $3\%$ at 820 MHz and less than $2\%$ at 600 MHz. Within the uncertainties, our determination of the galactic spectral index is practically unaffected by the correction for polarization.

Since the overall error budget of the sky temperatures measured by TRIS at 600 MHz, that is 66 mK(systematic)$+$18 mK (statistical), is definitely smaller than those reported in previous measurements at the same frequency, our data have been used to discuss the zero levels of the sky maps at 150, 408, 820 and 1420 MHz in literature. Concerning the 408 MHz survey, limiting our attention to the patch of sky corresponding to the region observed by TRIS, we suggest a correction of the base-level of $(+3.9\pm 0.6)$K.

\end{abstract}





\section{Introduction}

Measurements of brightness temperature and frequency spectrum of the galactic radio emission are of primary importance for astrophysics. There are tight links between the radio signal and the cosmic ray electrons which moving through the interstellar magnetic field produce synchrotron radiation. A detailed knowledge of the galactic radio emission is also important for observational cosmology. The galactic radio emission is a foreground which hampers the search and detection of spectral distortions, anisotropies and residual polarization of the Cosmic Microwave Background (CMB). Here we present and discuss results on the diffuse galactic emission at decimetric wavelengths obtained with TRIS, an experiment dedicated to the search for CMB spectral distortions, which included a subset of observations especially made to extract the galactic signal. The complete experiment and its cosmological results are presented in the two accompanying papers I \cite[]{zannoni08} and II \cite[]{gervasi2008a}. TRIS is a system of three absolute radiometers operating at nominal frequencies of 600, 820 and 2500 MHz respectively, installed at Campo Imperatore (Italy, latitude $=+42^{\circ}$). To minimize the level of radio-interferences at the observing site, the receivers were tuned at 600.5, 817.75 and 2427.75 MHz. The three antennas, wavelength-scaled pyramidal horns with the same beam ($HPBW_H\times HPBW_E = 23^{\circ} \times 18^{\circ}$) at the three frequencies, were aimed at the zenith with their E-planes tilted $7^\circ$ westward from the meridian. Along the right ascension, in a direction nearly parallel to the antennas' H-plane, the size of the beam was $23^{\circ}$. The radiometers have been used in two different operation modes: (1) in drift-scan mode to measure profiles of the variations of the sky temperature versus the right ascension $\alpha$, along a circle of constant declination $\delta = + 42^{\circ}$ and (2) in absolute mode, to get the absolute temperature of the sky at various points along the same circle. This configuration guarantees that the same sky region can be observed with the same beam at different frequencies, so that the comparison of the signals is straightforward. Moreover, every 24 hours a complete profile ({\it drift scan profile}) of the antenna temperature $T_a$ versus $\alpha$ is obtained. After subtraction of the local (ground and atmospheric) contributions, data collected at night time at different epochs during a year are combined in a profile of $T_{sky}$, the sky brightness temperature, covering the full right ascension range at $\delta =+42^{\circ}$ \cite[]{zannoni08}. The sky signal is a superposition of isotropic components of extragalactic origin, that is CMB and unresolved extragalactic radio sources (UERS, see \cite[]{gervasi2008b}), and the anisotropic galactic signal:

\begin{equation}\label{totalpower}
T_{sky}(\nu,\alpha,\delta)=T_{Gal}(\nu,\alpha,\delta)+T_{uers}(\nu)+T_{CMB}(\nu)
\end{equation}

\noindent where $T_{Gal}\propto \nu^{-\beta_{Gal}}$. The undefined frequency dependence of $T_{CMB}$ which appears in Eq. (\ref{totalpower}) accounts for possible spectral distortions of the CMB. \par\noindent Variations of $T_{sky}$ \textit{vs} $\alpha$ are due only to $T_{Gal}$.  Analyzing these variations by means of the T-T plot technique proposed by \cite{turtle62} we obtain $\beta_{Gal}^{TT}$. Having $\beta_{Gal}^{TT}$ it is then possible to disentangle the components of $T_{sky}$ \cite[]{gervasi2008a} and get absolute values of $T_{Gal}$. An important remark is that the physical interpretation of the values of $\beta_{Gal}$ found by means of this technique is not always straightforward. In particular, near the galactic plane, the spectral index of the diffuse emission found in this way is really produced by a blend of synchrotron and free-free emission. A possible approach followed by some authors (e.g. \cite[]{platania1998}) to extract the thermal contribution near the galactic plane, in particular at frequencies higher than 1 GHz, is to use an HII catalog.

The outline of the paper is the following. In Section \ref{dataset} we will describe the data set we used in our analysis. Then we will show the results obtained using just genuine TRIS data in Section \ref{trisresults}, while a more general discussion, including also polarization effects, can be found in Section \ref{generalconsideration}, where our data set is fully exploited. The astrophysical issues, and possible hints for future experiments, will be discussed in Section \ref{discussion}, followed by a last section summarizing the results (Section \ref{conclusions}).

\section{The data set}\label{dataset}

In this section we describe the starting point of our analysis, that is the data collected during the TRIS observational campaigns in drift-scan mode and the sky profiles extracted from extended radio surveys at 150, 408, 820 and 1420 MHz.
%% In a manner similar to \objectname authors can provide links to dataset
%% hosted at participating data centers via the \dataset{} command.  The
%% second curly bracket argument is printed in the text while the first
%% parentheses argument serves as the valid data set identifier.  Large
%% lists of data set are best provided in a table (see Table 3 for an example).
%% Valid data set identifiers should be obtained from the data center that
%% is currently hosting the data.
%%
%% Note that AASTeX interprets everything between the curly braces in the
%% macro as regular text, so any special characters, e.g. "#" or "_," must be
%% preceded by a backslash. Otherwise, you will get a LaTeX error when you
%% compile your manuscript.  Special characters do not
%% need to be escaped in the optional, square-bracket argument.

%% In this section, we use  the \subsection command to set off
%% a subsection.  \footnote is used to insert a footnote to the text.


%% This section also includes several of the displayed math environments
%% mentioned in the Author Guide.

\subsection{TRIS: drift scans and absolute measurements}\label{driftscans}

From paper I \cite[]{zannoni08} we take the two absolute profiles of $T_{sky}$
versus $\alpha$ at $\delta =+ 42^{\circ}$ measured at 600 and 820
MHz respectively. The data collected by TRIS at 2500 MHz cover only small portions of the right
ascension circle observed at lower frequencies; moreover, in these
portions, variations of $T_{sky}$ \textit{vs} $\alpha$, necessary to
recognize the galactic contribution, are of the same order of the
accuracy of $T_{sky}$. They are less accurate than other data
in literature at the same frequency \cite[]{sironi1984}.

\subsection{Sky profiles from other radio surveys}\label{others}

In order to obtain a more reliable (and more complete) study of the spectral index variation across the sky at $\delta=+42^{\circ}$, we decided to use also the 408 MHz \cite[]{haslam82} and 1420 MHz \cite[]{reich1986} surveys, both of them in the destriped version produced by \cite{platania2003}. We have extracted from their maps a sky profile at the declination observed by TRIS, assuming an elliptical
$23^{\circ}\times 18^{\circ}$ beam, and taking into account the orientation of the E-plane with respect to the meridian. Since the T-T plot technique is not sensitive to zero offset uncertainties, we considered the extracted profiles affected just by uncertainty upon the temperature scale, whose values at 408 MHz and 1420 MHz are respectively $10\%$ and $5\%$. To have a wider frequency coverage, we used also the $5^{\circ}$ resolution map at 150 MHz prepared by \cite{landecker70} using their Southern Sky survey at 150 MHz and the 178 MHz survey by \cite{tb62}. In this case, the reported temperature scale uncertainty is $5\%$.

Since we used single-polarization radiometers, we decided to check the consistency of our results exploiting also the Leiden polarization survey at 610 and 820 MHz \cite[]{brouw76}. Finally, we compared our absolute sky profile at $\delta=+42^{\circ}$ with the corresponding one extracted from the Dwingeloo northern sky survey at 820 MHz \cite[]{berkhuijsen72}.

\subsection{The Unresolved Extra-Galactic Radio Sources}

The contribution of the the Unresolved Extra-Galactic Radio Sources (UERS) to the brightness temperature collected by our antennas can be
considered as an isotropic signal scaling with frequency as a power law with spectral index $\gamma_{uers} \sim -2.70$. A detailed study of this signal, presented in a dedicated paper (\cite[]{gervasi2008b} and references therein), allowed us to disentangle the components of the celestial signal and, finally, to recover the amplitude of the galactic radio emission.

\section{TRIS}\label{trisresults}

\subsection{Methods of extraction of the galactic spectral index}\label{par_ttplots}

\par Two methods can be used to extract the galactic spectral
index:

\par i) {\it differential method}: based on
the T-T plot method (\cite{turtle62}) it can be used when values
of $T_{sky}$ measured at different positions on the sky at two
frequencies are available, no matter if the zeros of the scales of
temperature at frequencies $\nu_1$ and $\nu_2$ are known. Combining
Eq.(\ref{totalpower}) written for two frequencies $\nu_1$ and $\nu_2$ we obtain:

\begin{eqnarray}\label{TTplot}
T_{sky}(\nu_2,\alpha,\delta)=T_{sky}(\nu_1,\alpha,\delta)\Big(\frac{\nu_2}{\nu_1}\Big)^{-\beta_{Gal}(\alpha,\delta;
 \nu_1,\nu_2)}+
T_{uers}(\nu_1)\Big[\Big(\frac{\nu_2}{\nu_1}\Big)^{-\gamma_{uers}(\nu_1,\nu_2)} - \\
\nonumber
\Big(\frac{\nu_2}{\nu_1}\Big)^{-\beta_{Gal}(\alpha,\delta;
\nu_1,\nu_2)}\Big] + \Big[T_{CMB}(\nu_2) -
T_{CMB}(\nu_1)\Big(\frac{\nu_2}{\nu_1}\Big)^{-\beta_{Gal}(\alpha,\delta;
 \nu_1,\nu_2)} \Big].
\end{eqnarray}

\noindent In a sky region where $\beta_{Gal}$ does not depend on position, a plot of
$T_{sky}(\alpha,\delta;\nu_2)$ \textit{versus}
$T_{sky}(\alpha,\delta;\nu_1)$ is a straight line of slope

\begin{equation}\label{betaTT}
m = \Big(\frac{\nu_2}{\nu_1}\Big)^{-\beta^{TT}_{Gal}(\alpha_0,\delta_0;
 \nu_1,\nu_2)}
\end{equation}

\noindent where $\alpha_0$ and $\delta_0$ are the values of
$\alpha$ and $\delta$ at the center of the region used to build the T-T plot. Conversely, if the spectral index is not constant, loops and V-shaped features appear in the T-T plot.
\par\noindent The spectral index representative
of the uniform region centered on $\alpha_0$ and $\delta_0$ is
immediately obtained from $m$. Let's call it {\it average} or
$\beta^{TT}_{Gal}(\alpha_0,\delta_0;
 \nu_1,\nu_2)$  spectral index. It depends only on variations of
$T_{sky}$ and is independent on the zero level of the scales of
temperature used at $\nu_1$ and $\nu_2$.
\par\noindent

\par ii) {\it direct method}: when the
absolute values of $T_{Gal}(\alpha,\delta,\nu)$ at a given
position ($\alpha,\delta$) are known at two frequencies we can
write:
\begin{equation}\label{localindex}
\beta^{loc}_{Gal}(\alpha,\delta;\nu_1,\nu_2) =
-\frac{\log[{T_{Gal}(\alpha,\delta,\nu_1)/T_{Gal}(\alpha,\delta,\nu_2)}]}{\log{[\nu_1/\nu_2]}}
\end{equation}
\par\noindent Let's call it {\it local} spectral index.

Differential and direct methods, whose advantages and limitations have been described in \cite[]{lawson87}, are affected by systematics in completely orthogonal ways. The local spectral index is prone to the effect of wrong zero levels (see Eq.\ref{localindex}), but once we are confident that our absolute measurements are corrected for systematics, we can use this technique over the complete range of right ascension, regardless the nature of the physical properties of the observed regions. Conversely, $\beta_{Gal}^{TT}$ is not affected by temperature offsets, but it is not well determined over all the sky because the T-T plot is effective whenever (a) $\beta_{Gal}$ is only weakly dependent on celestial coordinates thanks to the homogeneity of the observed region and (b) the intensity of the signal varies appreciably through space. Since these conditions are not met everywhere, the T-T plot method cannot be used to get a well definite value of the spectral index at every point in the sky.

In the case of TRIS, in the sky regions where the T-T plot technique provides reliable results, we show that $\beta_{Gal}^{TT}$, which intrinsically represents a spatially averaged value of the spectral index, coincides with the local value of $\beta_{Gal}$ (Eq.\ref{localindex}). This is an important cross-check since, if the results are consistent within the uncertainties, then we can exclude undetected systematics affecting the zero levels of our absolute temperature scans, as we will show in Section \ref{Tris3}. If this is true, we may overcome the limitations of the T-T plot and assign a value of the local spectral index point by point in the sky.

\subsection{TRIS results}

\subsubsection{Spectral index from T-T plots}\label{SITT}

\begin{figure}
\begin{center}
\includegraphics[scale=0.7,angle=90]{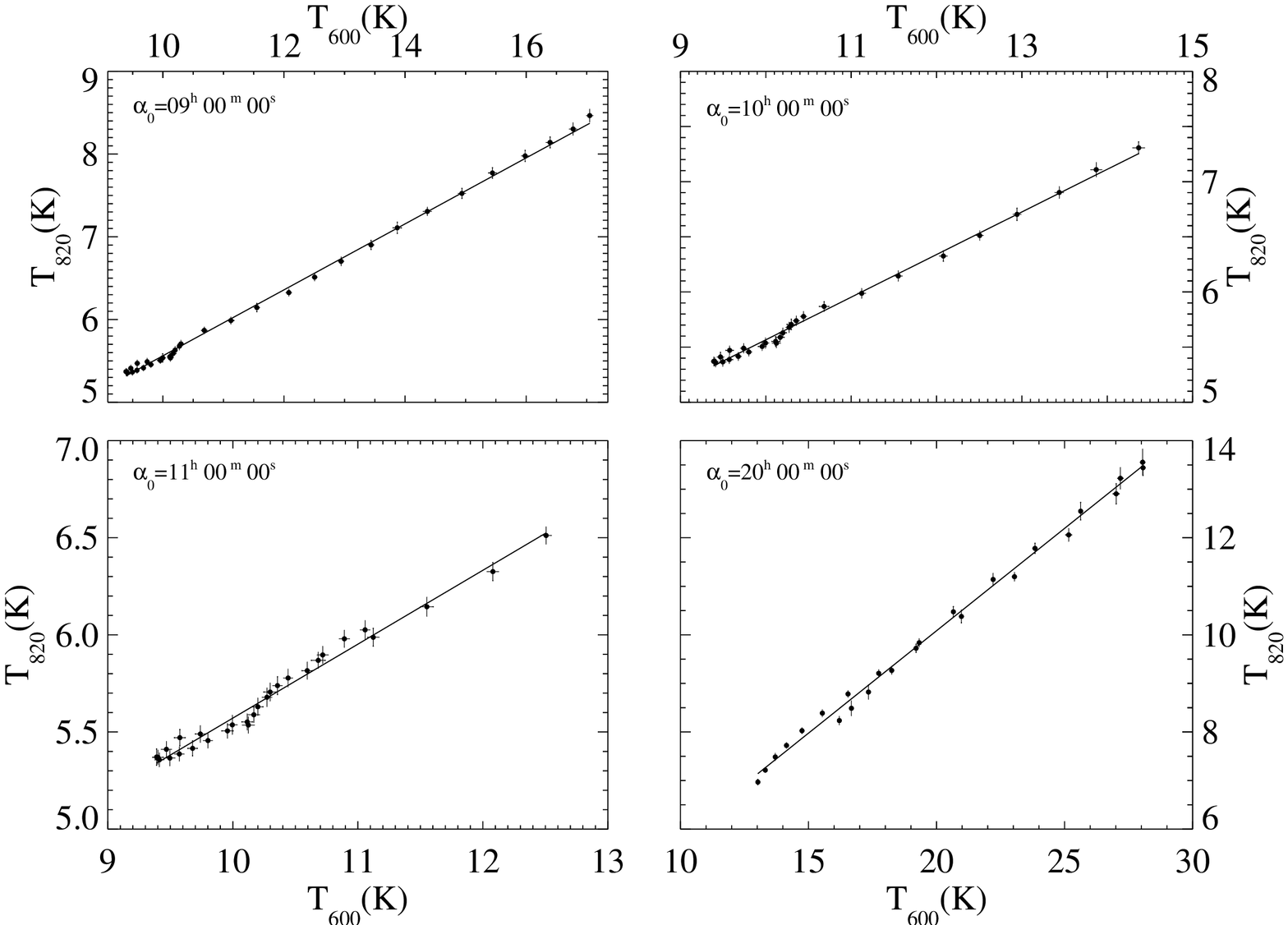}\caption{Examples of T-T plots of the sky regions we used to disentangle the astrophysical components of the celestial signal. In the upper left corner of each of the four panels the central value of the right ascension is reported.}\label{fig1}
\end{center}
\end{figure}

We apply the differential method to the pair of drift scans at 600
and 820 MHz measured by TRIS. As expected (see for instance
\cite{sironi1974}) the pattern produced by the complete set of $T_{sky}(\alpha)$ in the range
$0^h \leq \alpha \leq 24^h$ is far from a straight line
because $\beta_{Gal}$ varies with $\alpha$. There are however two
right ascension intervals along the circle at declination
$+42^{\circ}$ where straight lines are clearly defined and
$\beta_{Gal}^{TT}$ can be obtained (see Figure \ref{fig1}). The first
region covers the right ascension interval $(09^h \leq \alpha_0 \leq
11^h)$: it is characterized by a minimum of the sky brightness
temperature and signals come from the Galactic halo. The
second region is at $(19^h \leq \alpha_0 \leq 21^h)$. Here the sky
temperature reaches a maximum and the signal comes essentially from the
galactic disk.

In Table \ref{tab1} we show the values of $\beta^{TT}_{Gal}$ measured in
these regions and the width $\Delta \alpha$ of the right ascension
interval used to build the T-T plot. Lower limits on
$\Delta \alpha$ are set by the beam size of TRIS antennas and by
the statistical significance we want to assign to the parameters
extracted from the linear fit (for two parameters,  slope and
intercept, the degrees of freedom (\textit{dof}) are
$(2\Delta \alpha /FWHM) -2$). An upper limit on $\Delta \alpha$ is
set by the appearance in the plot of deviations from a straight
line, due the fact that we are considering a region wider than the homogeneity
scale of the interstellar medium.
We accepted fits with a chi-square probability Q (as defined
in \cite[]{press1992}) greater than 0.1. The effective values of Q
and $\chi^2/\textit{dof}$ are also shown in table \ref{tab1}
together with $\beta^{TT,robust}_{Gal}$ an estimate of the
galactic spectral index obtained using  a robust straight line
fitting function \cite[]{hoaglin1983}. The agreement within the uncertainties of
$\beta^{TT}_{Gal}$ and $\beta^{TT,robust}_{Gal}$ shows that the
selection criteria of the areas we use are properly chosen.

\subsubsection{Galactic Temperature and the Local Spectral Index} \label{Tris2}

\begin{figure}
\begin{center}
\includegraphics[scale=0.5]{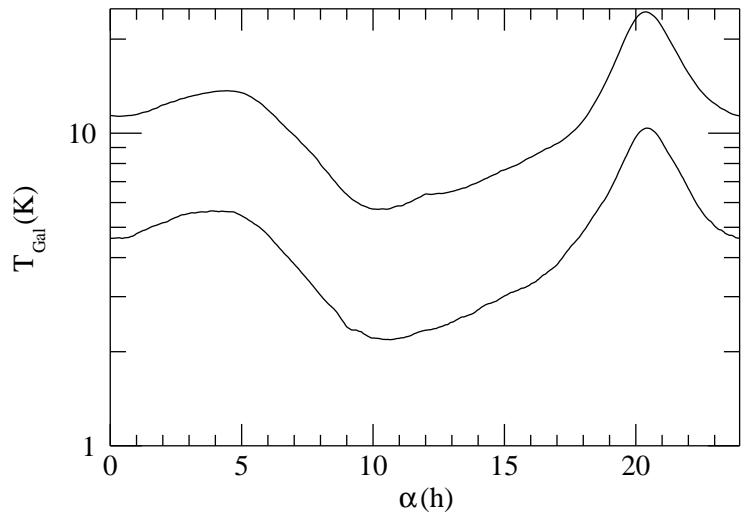}

\caption{The brightness temperature of the Galaxy at 600 (upper curve) and 820 MHz (lower curve) after the separation of the celestial signal
into a primordial (CMB), extra-galactic and galactic contributions. The correction for polarization based on \cite{brouw76} and
\cite{spoelstra84} is also taken into account. Some of the data plotted here are reported in Tab.\ref{tab2} with the corresponding uncertainties.}\label{fig2}

\end{center}
\end{figure}

\begin{figure}
\begin{center}
  \includegraphics[scale=0.8]{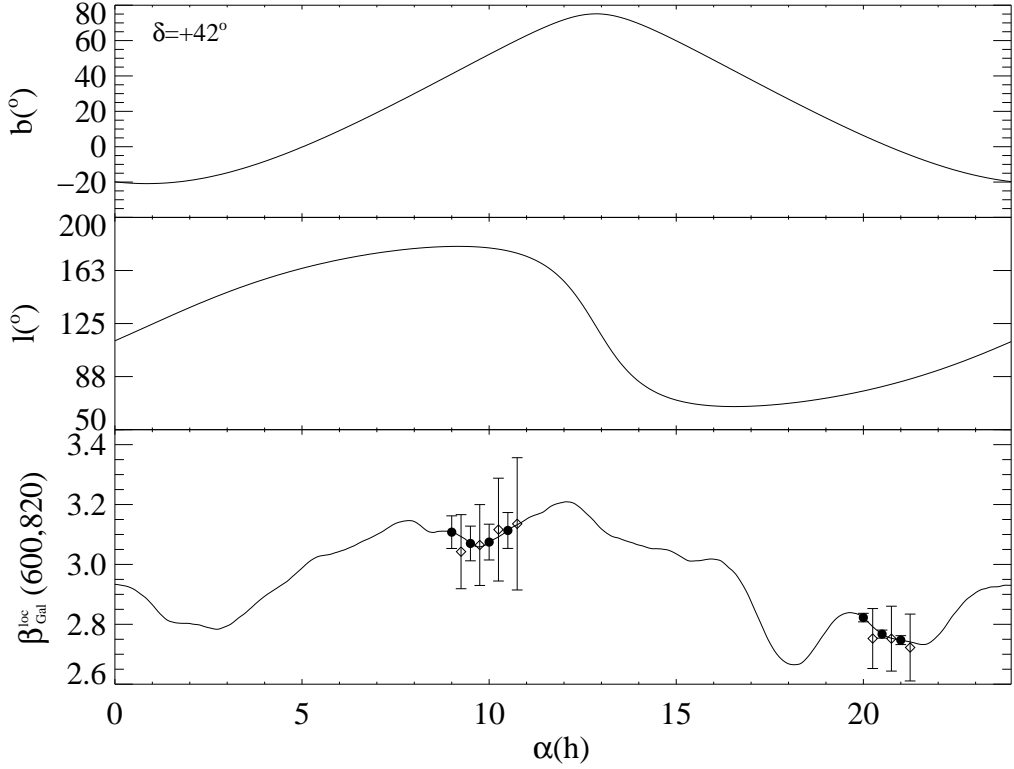}\\

    \caption{In the bottom panel: the local spectral index $\beta$ at $\delta=+42^\circ$ is shown in detail along the full right ascension range. We show also the values of $\beta_{Gal}^{TT}$ in the region where it has been accurately determined (open diamonds). In the same regions we plot the values of $\beta_{Gal}^{loc}$ (filled circles) accompanied by the corresponding uncertainties. The upper and central panels show the galactic latitude and longitude of the points along the circle at constant declination where TRIS observations were made.}\label{fig3}

\end{center}
\end{figure}

The above values of $\beta^{TT}_{Gal}$ cover only part of the full right
ascension circle at $\delta = +42^{\circ}$ explored by TRIS.
 However, combined with absolute measurement of $T_{sky}$ made by
TRIS, they are sufficient
to disentangle the components of $T_{sky}$ and get absolute
values of $T_{Gal}$ at 600 and 820 MHz at every point along the
circle at $\delta = +42^{\circ}$. This is done in paper II
\cite[]{gervasi2008a}. After corrections for polarization
effects (see Section \ref{par_polarization_effects}) the resulting values of
$T_{Gal}$ are used to get by Eq.(\ref{localindex}) the {\it local}
spectral index between 600 and 820 MHz. In Table \ref{tab2} and
Figure \ref{fig2} and \ref{fig3} we show values of $T_{Gal}$
corrected for polarization and values of $\beta^{loc,pol}_{Gal}$ and $\beta^{loc}_{Gal}$,
respectively the local spectral index corrected and not
corrected for polarization effects. The values of $T_{Gal}$
corrected for polarization are used  to get, by the T-T plot
method, values of $\beta^{TT,pol}_{Gal}$ corrected for
polarization. They are shown in Table \ref{tab3}. As expected the
smoothing effects of the large beam of the TRIS antennas make
statistically negligible the differences between the values of the
spectral index calculated including or not the polarization
effects.

\subsection{Constraints on TRIS systematic uncertainties}
\label{Tris3}
\par When both available, {\it local} and  {\it average} values of the galactic spectral
index coincide within the uncertainties (see Table \ref{tab1}, \ref{tab2} and bottom panel in Figure \ref{fig3}).
It means that the evaluations of the zero levels of
the scales of temperature made in paper I are correct and the systematic uncertainties $|\Delta T_{sys600}|\leq 66 ~mK$ and $|\Delta T_{sys820}|\leq
659 ~mK$ we associated to the absolute values of $T_{sky}$ in paper I are
safe. We cannot however exclude that these bars are overestimated.
They were in fact set by laboratory measurements of the
losses of all the front end components of the TRIS  radiometers.

\par These uncertainties can be further constrained if we exclude unphysical
situations they may imply. We can write

\begin{equation}\label{local_beta_sys}
\beta^{local}_{Gal}(\alpha;\Delta T_{sys600},\Delta T_{sys820} ) =
log\Big [ \frac{T_{Gal}(\nu_1;\alpha) + \Delta T_{sys600}
}{T_{Gal}(\nu_2;\alpha) + \Delta T_{sys820} }  \Big ]\Big /
log\Big( \frac{\nu_2}{\nu_1} \Big )
\end{equation}

\noindent and

\begin{equation}\label{discrepancy}
d(\alpha;\Delta T_{sys600},\Delta
T_{sys820})=\frac{|\beta^{local}_{Gal}(\alpha;\Delta
T_{sys600},\Delta T_{sys820}) - \beta
^{TT}_{Gal}(\alpha)|}{\sigma_{\beta^{TT}_{Gal}}(\alpha)+\sigma_{\beta^{loc}_{Gal}}(\alpha)},
\end{equation}
\par\noindent where $d$ is the difference between the {\it local}
and the {\it average} spectral index, in standard deviation units.
We then let the two $\Delta T_{sys}$ vary in such a way that $d
\leq 2$, while the spectral index stay inside an {\it a priori}
defined range of physical values. This is done using the TRIS data
collected between $\alpha = 09^h$ and $\alpha = 11^h$, where
$T_{Gal}$ is dominated by the synchrotron emission of the Cosmic
Ray Electrons and $\beta_{Gal} \cong \beta_{syn}$ (see Section
\ref{par_spectral_index}).

\par Measurements of the spectrum of the cosmic ray electrons
(see for example \cite[]{agrinier1964}; \cite[]{hartmann1977};
\cite[]{nishimura1990}; \cite[]{aguilar2002} and
\cite[]{grimani2002}) and the diffusion process of the electrons in
the Galaxy (e.g. \cite[]{longair1994} and references therein)
allow us to assume that, between 600 and 820 MHz, in the
direction of the galactic halo, a safe {\it a priori} condition is
($2.6 \leq \beta_{syn} \leq 3.6$). This choice constrains $\Delta
T_{sys820}$ inside the range (-300 mK, +430 mK). No constraints
are obtained at 600 MHz, nor using TRIS data from other regions of
sky, i.e. the galactic disk, where the presence of the thermal bremsstrahlung makes {\it a
priori} choices less reliable.

\section{General analysis}\label{generalconsideration}

\subsection{Correction for polarization effects}\label{par_polarization_effects}

\begin{figure}
\begin{center}
    \includegraphics[scale=0.4]{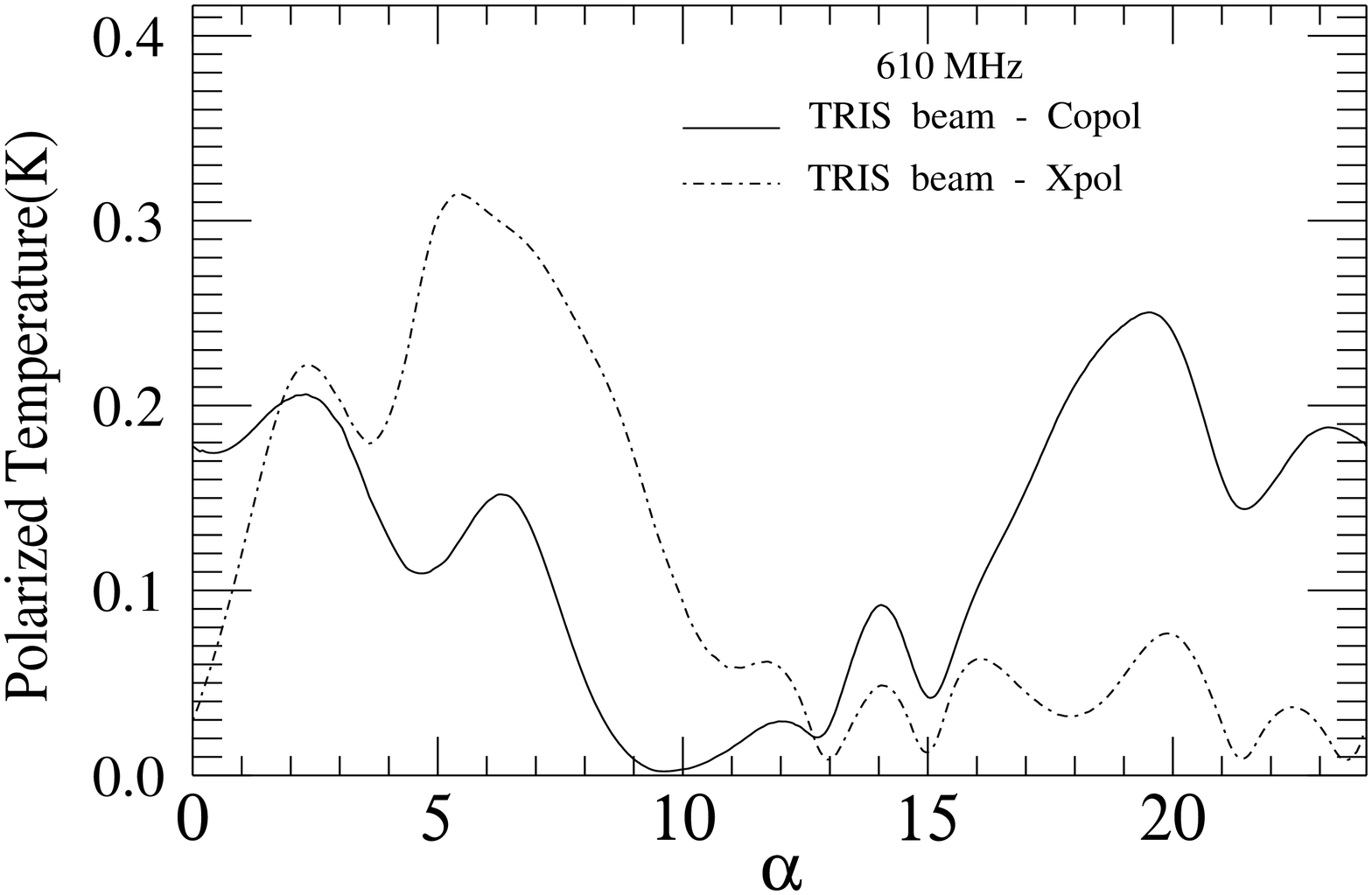}\\ \includegraphics[scale=0.4]{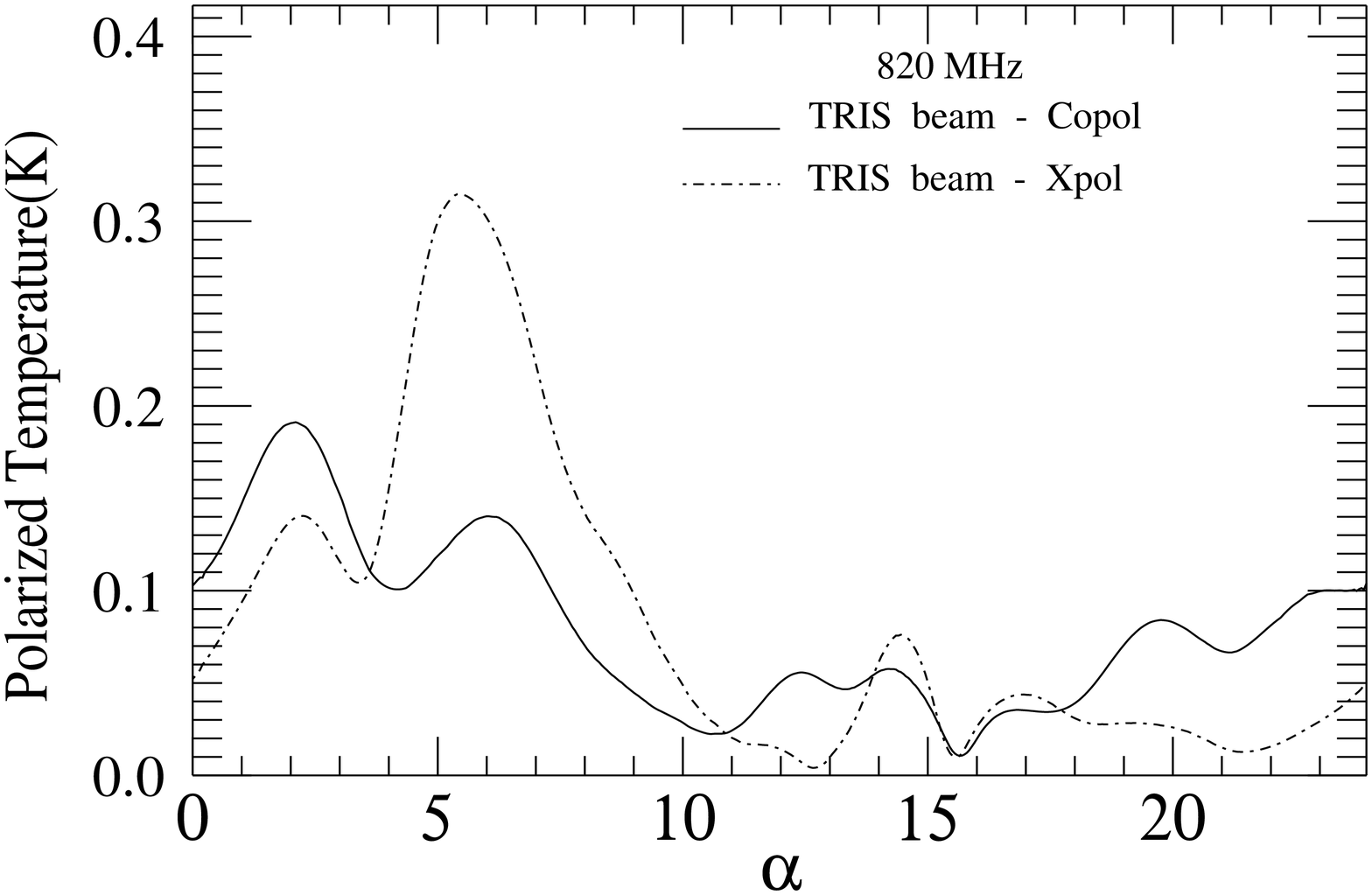}\\
    \caption{Upper panel: The co-polar and cross-polar temperature (K) \textit{vs} right
ascension extracted from Brouw and Spoelstra 610 MHz map \cite[]{brouw76} with respect to the TRIS antennas' E-plane and diluted on the
$23^{\circ}\times 18^{\circ}$  beam. Lower Panel: the same as upper panel, at 820 MHz.}\label{fig4}
\end{center}
\end{figure}

In order to evaluate polarization effects on our analysis
(particularly on the determination of the spectral index), we
corrected TRIS data with a contribution calculated from the
surveys by Brouw and Spoelstra \footnote{The original maps however
have not been our starting point, but rather we used a set of
Stokes' Q and U maps at $7^{\circ}$ resolution prepared by
E.Carretti (in the framework of the SPOrt program \cite[]{cortiglioni04}) starting from the original
data.} at 610 and 820 MHz (see \cite[]{brouw76} and the discussion
in \cite[]{spoelstra84}) convolved with the TRIS beam,
knowing the orientation of the polarization plane of TRIS antennas
point by point in the sky. In this way we have been able to
project the polarized signal into copolar ($T_{copol}$) and
crosspolar ($T_{Xpol}$) components at the two frequencies (see
Figure \ref{fig4}). Then we have rescaled the 610 MHz sky
profile to 600 MHz by means of the spectral index derived from
TRIS data alone. Finally, we obtained the antenna temperature of the
sky corrected for polarization:

\begin{equation}\label{correction_polarization}
T_{sky}(\alpha; \nu) = T_{TRIS}(\alpha; \nu) - T_{copol}(\alpha; \nu) +  \sqrt{T_{copol}^2(\alpha; \nu) + T_{Xpol}^2(\alpha; \nu)}.
\end{equation}

\noindent These corrected sky profiles can be used to evaluate the spectral index by means of T-T plots (Eq.\ref{TTplot}), or using
Eq.\ref{localindex}, since we have absolute temperature profiles. The results, concerning the local galactic spectral index, are presented in Table \ref{tab2}, together with the corrected temperatures, and shown in Figure \ref{fig5}. A comparison between the results reported in
Table \ref{tab1} and Table \ref{tab3} (obtained after polarization correction), gives an estimate of the effect of polarization on the spectral
index ($\Delta \beta_{Gal}^{TT}$) for an experiment with $FWHM \sim 20^{\circ}$: $\Delta \beta_{Gal}^{TT} \cong 0.0$ on the galactic disk towards the center direction; $\Delta \beta_{Gal}^{TT} \cong -0.1$ towards the halo, where, anyway, the effect is comparable with the uncertainties.

\noindent TRIS data, after correction for polarization, have been used also in combination with sky profiles at $\delta=+42^{\circ}$ extracted from surveys at 150, 408 and 1420 MHz, always using the T-T plot technique, described in Section \ref{par_ttplots}. The criteria used to single out acceptable values of the fitting parameters are the same described in section \ref{SITT}. The results are shown from Table \ref{tab4} to Table \ref{tab7}. A summary of the results obtained in four distinct sky positions, namely the galactic halo ($\alpha = 09^h 00^m$, $\alpha = 10^h 00^m$ and $\alpha = 11^h 00^m$ at $\delta=+42^{\circ}$), and the galactic disk ($\alpha = 20^h 24^m$, at the same declination) are shown in Table \ref{tab8}.

\subsection{A new assessment of the zero level of continuum surveys}\label{par_zero_maps}

The temperatures measured by TRIS at 600 MHz have accuracies of (66 (systematics) + 18 (statistics))~mK
on $T_{sky}$ and $\sim 70$ ~mK on $T_{Gal}$ (\cite{zannoni08}
and \cite{gervasi2008a}). They are definitely more accurate than
previous measurements in literature at the same frequency (\cite[]{sironi1990} and references therein). This accuracy has been
exploited to revise and sometimes improve the accuracy of the zero
level of the maps of the radio continuum at 150, 408, 610, 820 and
1420 MHz in literature.
This is done comparing profiles extracted from these maps
with profiles obtained at the same frequency by extrapolation
of the of astrophysical signals (Galaxy,UERS and CMB) contributing to the $T_{sky}$ measured by TRIS at 600 MHz.

The comparison is made at ($\alpha_m = 10^h 00^m$,
$\delta_m=+42^{\circ}$), where $T_{sky}(600)$ reaches a minimum,
$\beta_{Gal}^{TT}$ and $\beta^{loc}_{Gal}$ are both well defined and $T_{sky}$
has a weak dependence on the celestial coordinates. Here minor antenna
pointing errors, projection effects, and discrepancies between the
TRIS beam and the synthesized beam assumed in making map convolutions,
have a negligible impact upon the determination of the
antenna temperature at TRIS resolution.

The starting point of this analysis is the first line in Table \ref{tab9}, where we recall the amplitude of the three signals contributing to the sky temperature measured by TRIS at 600 MHz at ($\alpha_m$, $\delta_m$). Then, knowing the spectral index determined by means of T-T plots, we evaluated the galactic contribution at the frequencies of the different surveys, from 150 to 1420 MHz. These values are reported in the second column of Table \ref{tab9}, where the quoted uncertainties are obtained propagating both the uncertainty on the spectral index obtained between 600 MHz and the considered frequency (see Table \ref{tab8} for a summary) and on the amplitude of the galactic signal at 600 MHz. The third column in Table \ref{tab9} recalls the estimates of the extragalactic contribution calculated by \cite[]{gervasi2008b}. The fourth column is the brightness temperature of the CMB measured at 600 MHz, accompanied by its overall uncertainty. All these three contributions are then combined in the fifth column, $T_{eval}^{(\alpha_m,\delta_m)}$, which, in turn, is compared with the temperature of the sky profiles extracted from the surveys, reported in the sixth column, with associated zero-level systematics quoted in literature. The difference between the two, $T_{eval}^{(\alpha_m,\delta_m)} - T_{map}^{(\alpha_m,\delta_m)}$, reported in the last column, gives the correction we suggest to apply to the zero levels of the maps used throughout this work.

At 150 MHz the zero level adjustment is large, but comparable to
the accuracy of the old zero level. At 408 and 820 MHz adjustments
accuracies are significant and cannot be ignored. At 1420 MHz no
base level correction is necessary but the accuracy is improved by
a factor 3.

\subsection{Galactic spectral index}\label{par_spectral_index}

In spite of the limited sky coverage of TRIS  and the coarse
angular resolution of its antennas, the local spectral index and
its variations can be tracked along the complete right ascension
circle at $\delta = +42^{\circ}$. It brings information about
different galactic regions going from the halo to the disk. To point out this fact
the values of the local spectral index
$\beta^{loc}_{Gal}(600,820)$ along the circle at $\delta = +
42^{\circ}$ have been plotted in Figure \ref{fig3} \textit{vs} right ascension, together with
the corresponding galactic coordinates. In
agreement with previous observations (e.g. \cite{sironi1974} and
\cite{webster1974}) a steepening of the spectrum is observed when
the line of sight moves from low to high galactic latitudes.
Dependence on galactic longitude, which partially masks the
latitude dependence, is also visible. This is not surprising
because along the circle at $\delta = + 42^{\circ}$ the
galactic disk is crossed twice, in regions where the relative
weights of thermal and non-thermal emission are different. Moreover,
extended structures coming out from the disk, such as the galactic
loop III \cite[]{berkhuijsen71}, visible on the all sky maps, play
an important role in the determination of the radio continuum
spectral index \cite[]{lawson87}.

\subsubsection{The galactic halo}\label{galactichalo}

From T-T plots centered in regions belonging to the galactic halo, where the synchrotron is the dominant radiative process, we found that between 150 MHz and 600 MHz, as a consequence of a knee in the energy distribution of cosmic rays electrons, the brightness temperature spectral index, even if in some cases it is poorly determined, changes from $2.3\div 2.5$ to
$2.8\div 3.0$, as we can see in Table \ref{tab8}
and Figure \ref{fig6}. This behavior, observed in the past (e.g. \cite{sironi1974} and
\cite{webster1974}), and confirmed more recently
(\cite{roger1999}), can be considered well understood (see for instance
\cite{longair1994}). This feature is also observed looking at the galactic temperature \textit{vs} frequency, as shown in Figure \ref{fig7}.

Since in the halo HII regions are absent, synchrotron is the
dominant radiative process and $\beta_{Gal} \sim \beta_{syn}
=(a+3)/2$ where ~$a$~ is the spectral index of the energy spectrum
$N(E) dE = K E^{-a} dE$ of the Cosmic Ray Electrons (CRE)
responsible for the synchrotron emission.  The increase of
$\beta_{syn}$ with $\nu$ is probably a consequence of a knee, at
$E_k \sim 1$ GeV, in the CRE energy spectrum: with $a=a_1$ at
$E<E_k$ and $a=a_2$ at $E>E_k$. No matter how sharp is the knee,
the change of $\beta_{syn}$ from $\beta_1 = (a_1 + 3)/2$ at low
frequencies to $\beta_2 = (a_2 + 3)/2$ at high frequencies takes
place gradually. Following \cite{sironi1969} we can however match
$E_k$ to a frequency $\nu_{ex}(E_k,B_\bot)=f(a_1,a_2)\nu_c(E_k,B_\bot)$
where $f$ is a function of CRE spectral indices, while $\nu_c = A B_\bot E^2$ (where A is a constant)
is the well known {\it critical frequency} which links the frequency of synchrotron
emission, the energy $E$ of the radiating electron and the component
$B_\bot$ of the magnetic field $B$ orthogonal to the line of sight
(e.g \cite{ginzburgsyrovatskii1965}).

 In Figure \ref{fig8} we show the contour lines of $\nu_{ex}$ in the ($E_k$,$B_\bot$) plane for $a_1=1.7$ (consistent with $\beta_{syn}\simeq 2.3$, and contained in the range 1.6$\div$ 1.8 suggested by \cite{strong2000}) and for two values of $a_2$, that is 2.6 and 3.0, corresponding respectively to $\beta_{syn1}=2.8$ and $\beta_{syn2}=3.0$, the allowed interval of values for $\nu\geq$600 MHz according to our analysis. Steeper spectra, reaching also $\beta_{syn}\cong 3.3$, can be produced in a scenario in which magnetic inhomogeneities are considered \cite[]{cavallo77}, or in presence of a further steepening of the CRE energy spectrum. In the latter case, the effect on the radio spectrum should be relevant especially at tens of GHz, since $a=3.6\pm 0.2$ is found for energies $E\gtrsim$6 GeV \cite[]{boezio2000}.

From the results obtained in the present work, based only on data coming from radio-astronomy, we can state that towards the galactic halo $\nu_{ex}<$600 MHz, as we can see in Figure \ref{fig6} and \ref{fig7}, and Table \ref{tab8}. If we take into account also the spectral index map presented in \cite[]{roger1999}, where no synchrotron knee is detected between 22 and 408 MHz in the region observed by TRIS, a tighter bound is 408MHz$<\nu_{ex}<$600 MHz. This
constraint on $\nu_{ex}$ singles out a region of allowed values for $E_k$ and $B_\bot$ (see Figure \ref{fig8}), but the degeneracy cannot be broken without a further relationship between the knee energy and the galactic magnetic field provided by the modelling of CRE's diffusion-loss process with additional data-sets (see for example \cite[]{strong2000}).

\subsubsection{The galactic disk}

Moving towards the galactic disk (see last column in Table \ref{tab8}), the T-T plots between one of the low frequency sky profiles (408, 600, 820 MHz) and the 1420 MHz profile show that the radio spectrum flattens from $\sim 2.8$ to $\sim 2.5$, probably because the weight of HII regions, especially at 1420 MHz, is no longer negligible. In this case $\beta$ has to be interpreted as an effective spectral index ($\beta_{eff}$), which is properly the result of a blend of thermal (free-free) and synchrotron emission. If we have the galactic temperature at two frequencies $\nu$ and $\nu_0$, then

\begin{equation}\label{ff_separation}
T_{Gal}(\nu)= T_{syn}(\nu/\nu_0)^{-\beta_{syn}}+T_{ff}(\nu/\nu_0)^{-2.1}
\end{equation}

\noindent while $T_{Gal}(\nu)/T_{Gal}(\nu_0)=(\nu/\nu_0)^{-\beta_{eff}}$. Defining $w=T_{ff}/T_{syn}$, the ratio of the free-free and
synchrotron contributions at $\nu_0$, we obtain

\begin{equation}\label{peso_ff_syn}
\beta_{eff} = log\Big [ \frac{(\nu/\nu_0)^{-\beta_{syn}}+w(\nu/\nu_0)^{-2.1}}{1+w}  \Big ]\Big / log\Big( \frac{\nu_0}{\nu} \Big )
\end{equation}

\noindent which tell us that $\beta_{eff}<\beta_{syn}$, since $w$ is always positive and $\beta_{syn}>2.1$.

After the adjustment of base-levels described in Section \ref{par_zero_maps}, we used the temperatures in correspondence of the peak of the galactic signal at 408, 600.5 (TRIS), 817.5 (TRIS), 820 and 1420 MHz to fit the amplitudes of the galactic components, $T_{syn}$ and $T_{ff}$ and the value of $\beta_{syn}$. We also imposed that the condition $T_{ff}+T_{syn}=T_{Gal}(\nu_0)$ was fulfilled within the uncertainties, $T_{Gal}(\nu_0)$ being the measurement we have obtained with TRIS at the frequency $\nu_0=$600.5 MHz. We obtained $\beta_{syn}=2.95\pm 0.08$ and $w=0.13\pm 0.03$. The results of the fit together with the data are shown in Figure \ref{fig9}. This means that in this region of the sky (the Cygnus region) we can assign to the thermal component a weight around 11$\%$ of the total emission ($T_{ff}/T_{Gal}(\nu_0)$) at 600 MHz. With this value we obtain an effective spectral index $\beta_{eff}= 2.84 \pm 0.04$ between 600 and 820 MHz, which is fully consistent with the value found by means of T-T plots, $2.8\pm 0.1$. At 1420 MHz we can calculate a weight of the thermal emission around 21$\%$ (0.21$\pm$ 0.05), to be compared with the 40$\%$ quoted by \cite{reich1988b} (in the same way, their value at 408 MHz is between 11$\%$ and 16$\%$, while we have found 9$\pm$2$\%$). In this case a discrepancy may be ascribed to the fact that we used a very wide beam ($\sim$20$^{\circ}$FWHM, which intrinsically makes our experiment more sensitive to the more diffuse signals), or it may be simply accidental, since TRIS crossed the galactic disk in the center direction only
in one position. Moreover, we haven't considered a possible variation of the synchrotron spectral index between 408 and 600 MHz, but the
position of the point at 408 MHz in Figure \ref{fig9} suggests that $\beta_{syn}$ flattens at lower frequencies also in the disk region.

\subsection{Reconstruction of the 2500 MHz profile}

Having no reliable drift scan at 2500 MHz, we built a profile of the galactic component starting from the
Stockert \cite[]{reich1986} survey at 1420 MHz. This work is needed in order to extract the CMB signal from the sky temperature.

\noindent To obtain the galactic sky profile at 2500 MHz, we first convolve this map with the beam of the TRIS antennas in order to get the
synthetic drift scan at $\delta =+42^\circ$. Then  we have to exploit the data available from the analysis described in the previous sections and from paper II: (1) both the amplitude of the galactic signal at 600 and 820 MHz and (2) the galactic spectral index between the TRIS frequencies and the 1420 MHz survey in the two sky regions where the T-T plot gives reliable results. Using these quantities, we can extrapolate the pure galactic contribution from 820 to 1420 MHz knowing the spectral index $\beta_{Gal}^{TT}(820, 1420)$. In the regions where this extrapolation is possible, i.e. where $\beta_{Gal}^{TT}$ is well defined (that is in the two regions described in  Table \ref{tab6} and Table \ref{tab7}), we can single out a common offset which can be subtracted from the full scan at 1420 MHz. In this way, we obtain a galactic profile at 1420 MHz, to be further rescaled at 2500 MHz.

Since we need a full sky circle at $\delta=+42^\circ$ at 2500 MHz, in order to rescale the galactic contribution from 1420 to 2500 MHz, we have to use the spectral index $\beta_{Gal}^{loc}$ between 600 and 820 MHz, which is well defined everywhere in our right ascension range and, moreover, derived using only measured quantities. In this case, we are assuming that there are no changes in the slope of the galactic emission at frequencies higher than 820 MHz. This is a safe assumption for the signal coming from the halo (see Section \ref{galactichalo}), essentially synchrotron emission, while this extrapolation is more critical for regions close to disk, where the free-free is no longer negligible.
We could not use the $\beta_{Gal}^{loc}$ between 820 and 1420 MHz (which \textit{a priori} would be the natural choice), nor between 600 and 1420 MHz, since its uncertainty doesn't allow to recognize the shape of the galactic profile at 2500 MHz. This is due essentially to the fact that at high galactic latitudes the galactic signal at 1420 MHz is of the same order of magnitude of the uncertainty affecting the zero of the temperature scale at this frequency.
Following this procedure, we obtain a value of $T_{Gal}=91 \pm 93$ mK at $\alpha = 10^h 00^m$, while the peak value corresponding to the galactic disk is $471 \pm 93$ mK. Here we quote only the systematic uncertainty which is the dominant one.

\noindent The result obtained in this way has been used in paper II in order to subtract the galactic contribution from $T_{sky}$ at 2500 MHz, at galactic latitudes $b\gtrsim 20^\circ$ (which means one full beam over the galactic plane).

\section{Discussion}\label{discussion}

\subsection{Spectral coverage, sky coverage and polarization}

If we can choose two frequencies $\nu_1 < \nu_2$ in such a way that in this frequency interval there are no changes in the slope of the radio brightness power law, the spectral index is given in Eq.\ref{localindex}, provided that we have been able to disentangle the galactic contribution from the overall signal at the two frequencies. But where we cannot say \textit{a priori} if there's a knee in the radio spectrum, using the spectral index derived in this way to estimate the emission at a third frequency $\nu_3$, with $\nu_1 <\nu_3 < \nu_2$, we could be grossly wrong. This is the reason why we could not rely just on the 408 MHz and 1420 MHz surveys to disentangle the galactic emission, say, from few absolute data points at 600 and 820 MHz, but rather we needed the reconstruction of sky profiles across the full right ascension range at our frequencies in order to obtain a better characterization of the spectral index. Also a very rough analysis clearly shows that the relative weight of the components of the diffuse radiation is far from being constant. Therefore the \emph{sky coverage} should be considered very carefully in a future \emph{multi-frequency} experiment aimed at the study of the CMB at decimetric wavelengths, since the Position Difference Technique described in \cite[]{gervasi2008a}, which proved itself to be particularly effective in components separation, takes advantage exactly from the halo-disk modulation of the galactic signal.

The T-T plot technique allows us to assign a spectral index to the anisotropic (galactic) component of the celestial signal regardless its
nature, a step needed in order to separate this contribution from the isotropic ones, that is the monopole term of the CMB and the integrated effect of unresolved extra-galactic radio sources. At 600 and 820 MHz we used our sky profiles without any kind of correction based on external data-sets, since for foreground removal, according to the scheme described in paper II, we needed to single out just the signals as seen by our antennas. A second, straightforward comment, related to polarization issues, is that in principle the spectral index given in Table \ref{tab1}, even if looking at the galactic halo where synchrotron is the dominant mechanism, is not rigorously the true synchrotron spectral index, which instead is intimately linked to the degree of polarization of the radiation. Nevertheless, (1) this was what we needed for TRIS immediate purposes and (2) we checked that the correction for polarization produces an effect comparable to the uncertainty affecting the the determination of $\beta_{Gal}$ itself, probably due to beam depolarization. Therefore, in view of future experimental programs, we stress that a dual polarization system, although not mandatory, would allow (1) a better control of systematic effects, (2) the direct measurement of the full (polarized + unpolarized) sky temperature and, as an ancillary result, (3) it could produce an important advance in the knowledge of the diffuse galactic radio emission.

\subsection{The effect of baselevel corrections on the local spectral index}

We conclude this section with some further remarks concerning our analysis of the zero levels of continuum surveys used in this work. We
emphasize that, at least in the Northern Sky, where direct measurements have been carried out, the choice of the new zero levels we suggest in Section \ref{par_zero_maps} should be done in order to prevent baselevel effects in the determination of the local galactic spectral index. We checked the effect of this corrections on the local spectral index calculated using the sky profiles at 408 and 1420 MHz extracted from the destriped maps \cite[]{platania2003}. The effect is obviously a general steepening in the spectrum, since a higher temperature is associated to the 408 MHz profile, but while this effect is practically negligible around the Cygnus region (galactic disk), since its value here is $\Delta \beta \cong +0.04$, it is decisive around the minimum of the galactic signal at $\delta=+42^{\circ}$ (galactic halo), where $\Delta \beta \cong +0.2$. Generally speaking, the steepening will be more pronounced at high galactic latitudes, where the galactic emission becomes fainter and fainter and where the signal is comparable with the constant baselevel correction. A further investigation of this effect is not in the purposes of this paper, but nevertheless we suggest that it should be carried out in future works. We recall that the problem of the flattening of the radio spectrum between 408 and 1420 MHz at high galactic latitudes, even at an angular resolution of 2$^{\circ}$, has already been noticed and discussed by \cite{reich1988b}, on the basis of their spectral index map (whose absolute spectral index error is $\cong$ 0.1 \cite[]{reich1988a}), within the framework of a galactic wind model. Some years later, and from a different viewpoint, this problem has been noted by \cite{davies1996} studying a strip of sky centered at $\delta=+40^{\circ}$. In fact, using T-T plots, they have found a steepening of the spectral index going from the galactic plane towards the halo and, in the same work, they suggested also the existence of a baselevel residual $\sim$3.5 K in the 408 MHz survey ($\sim$0.15 K at 1420 MHz). The correction of the zero levels of these maps we have suggested in Table \ref{tab9} seems to solve the problems, but we think that a further analysis based on a better removal of baselevel effects and new observations are essential to clarify the situation.

\section{Conclusions} \label{conclusions}

The main purpose of the study we presented in this paper was to evaluate the galactic diffuse emission at decimetric wavelengths at $\delta=+42^\circ$. This work is necessary to extract the temperature of the Cosmic Microwave Background at frequencies
close to 1 GHz from TRIS absolute measurements. In fact, distortions of the blackbody CMB
spectrum are expected at decimetric wavelengths \cite[]{gervasi2008a}.

\noindent We extended however our analysis of the galactic signals and added
to our measurements data in literature covering the frequency
range 150 MHz - 1420 MHz.  We conclude that at angular resolutions
of $\sim 20^{\circ}$: i) the steepening of the galactic spectral
index from $\beta_{Gal} \simeq 2.3\div 2.5$ to $\beta_{Gal} \simeq 2.8 \div 3.0$ at
high galactic latitude is confirmed; ii) the change of slope is
maximum between 408 and 600 MHz; iii) this behavior is fully
compatible with the features of the Cosmic Ray Electron energy
spectrum and the intensity of the Galactic magnetic field; iv) in
the disk the signal is a mixture of synchrotron and thermal
emission from HII regions, with HII contribution to $T_{Gal}$
variable from 11$\%$ at 600 MHz to 21$\%$ at 1420 MHz; v) at this angular resolution the polarization
of the synchrotron radiation affects for less of 1$\%$ the spectral index values.

We have also shown, in section \ref{Tris3}, how a detailed knowledge of the spectral index can help in recognizing and eventually
controlling systematic uncertainties affecting absolute temperature scans.

\begin{figure}
\begin{center}
    \includegraphics[scale=0.5]{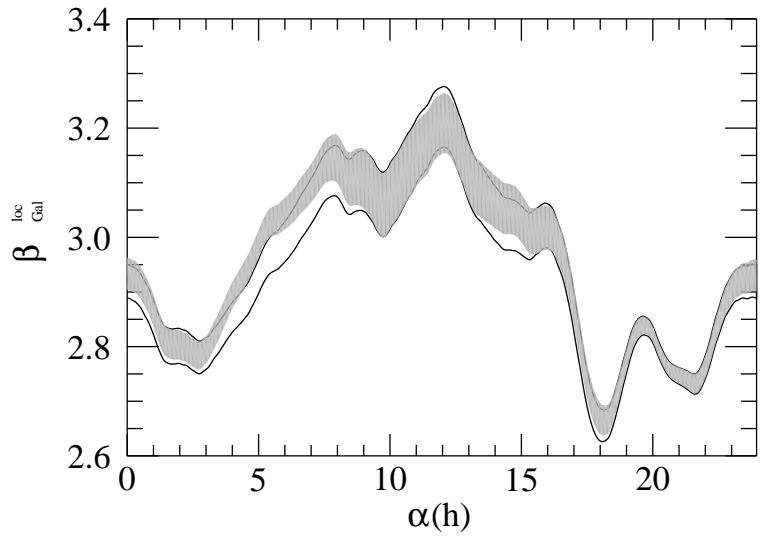}\\
    \caption{The local spectral index $\beta_{Gal}^{loc}$ between 600 MHz and 820 MHz (TRIS data) at $\delta=+42^{\circ}$. Grey band: data not corrected for polarization effects. Region enclosed by solid lines: data corrected for polarization. The width of the two bands represents the $\pm \sigma$ uncertainty. The two bands, which have a similar width, are almost everywhere superposed, except in the region $04^h \lesssim \alpha \lesssim 08^h$.}\label{fig5}
\end{center}
\end{figure}

\begin{figure}
\begin{center}
\includegraphics[scale=1.]{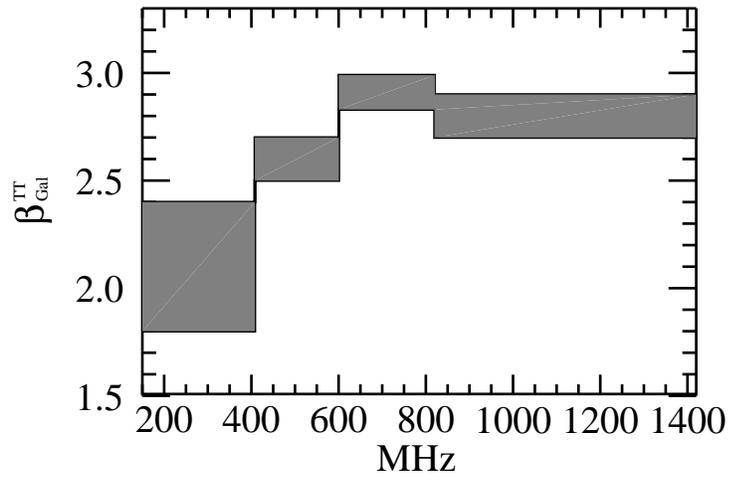}

\caption{The grey band represents the galactic spectral index derived from T-T plots, between the two frequencies which limit
each marked interval, in the halo region at $\alpha = 09^h 00^m$.}\label{fig6}

\end{center}
\end{figure}

\begin{figure}
\begin{center}
\includegraphics[scale=0.8]{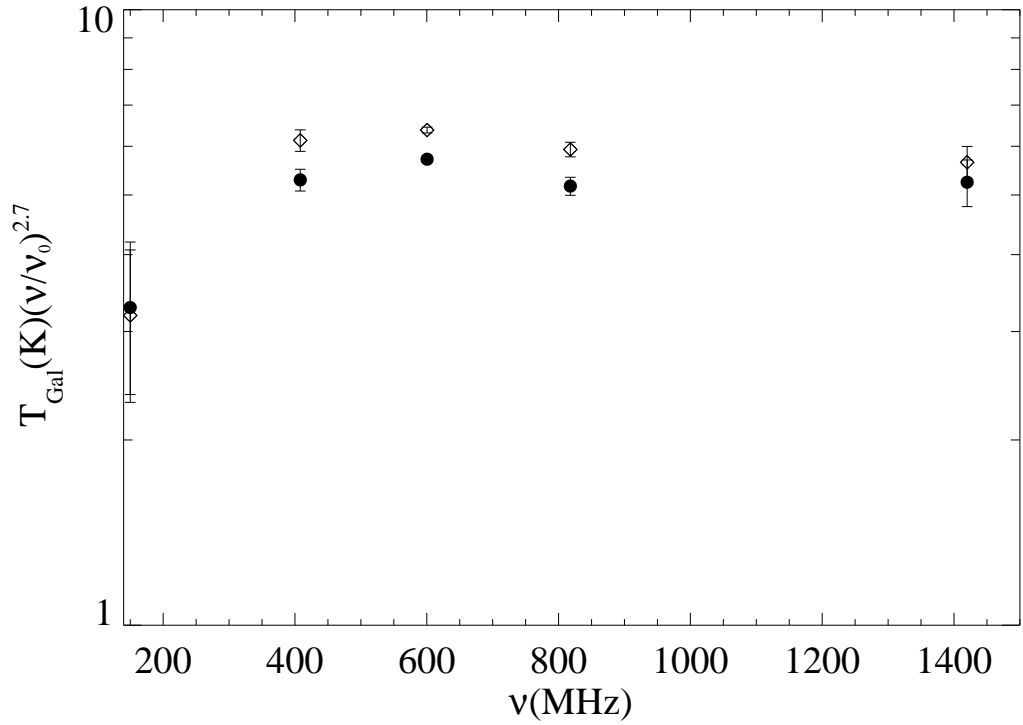}

\caption{The antenna temperature of the galactic emission (multiplied by a factor $(\nu (MHz)/600)^{2.7}$) at ($\alpha = 09^h00^m$,
$\delta=+42^{\circ}$) (diamonds) and ($\alpha = 10^h00^m$; $\delta=+42^{\circ}$) (filled circles). The temperature is plotted \textit{vs} frequency (MHz). Data points at 150, 408, 820 and 1420 MHz have been obtained after the zero level adjustment of the maps (see Sec.\ref{par_zero_maps}).}\label{fig7}

\end{center}
\end{figure}

\begin{figure}
\begin{center}
\includegraphics[scale=0.5]{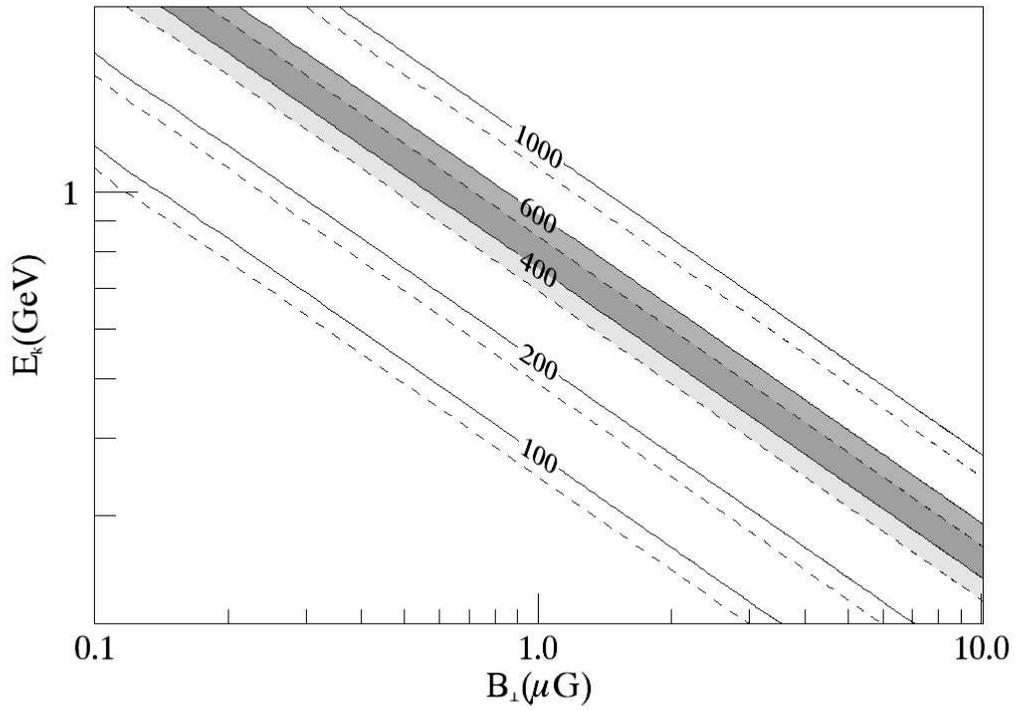}

\caption{Contour levels of $\nu_{ex}$(MHz) in the ($E_k$,$B_\bot$) plane. The two families of curves are referred to $a_2=3.0$ (dashed lines)
and $a_2=2.6$ (solid lines) with $a_1=1.7$ (see Sec.\ref{par_spectral_index} for explanation). The regions with 400$\leq \nu_{ex}(MHz)\leq $600
are shown as distinct (and partially superposed) grey bands for the two cases.}\label{fig8}

\end{center}
\end{figure}

\begin{figure}
\begin{center}
\includegraphics[scale=0.6]{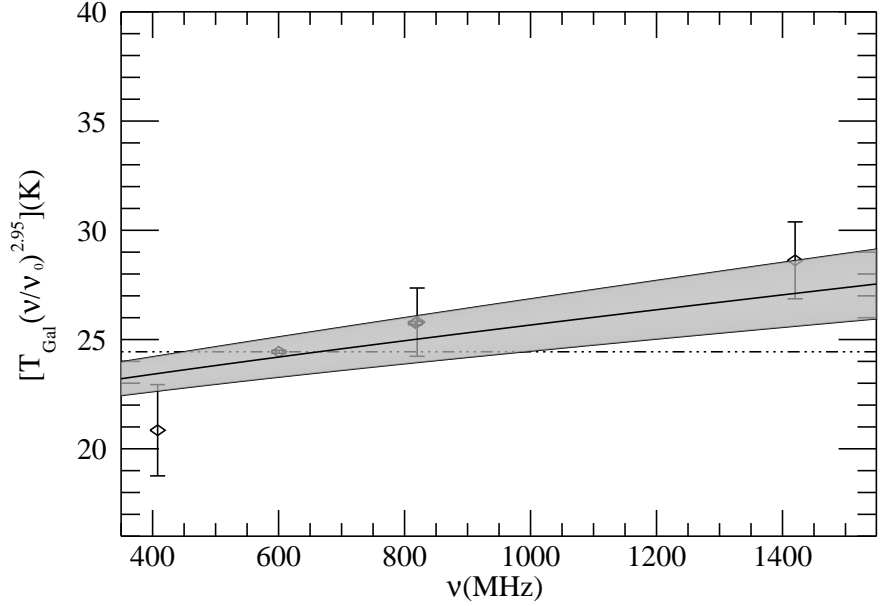}

\caption{Temperature of the Galaxy towards the Cygnus region ($\alpha = 20^h 24^m$, $\delta=+42^{\circ}$). The plot represents
$T_{Gal}(K)(\nu/\nu_0)^{2.95}$ \textit{vs} $\nu$ (MHz), with $\nu_0=600.5$ MHz. Diamonds: experimental points at 408, 600.5, 817.85, 820 and 1420 MHz. The dash-dotted line is the temperature at 600 MHz scaled at the different frequencies with a synchrotron spectral index $\beta_{syn}=2.95$.
The light grey band represents the result of the two components (thermal + non-thermal) fit. The error bar at 600.5 MHz is smaller than the
plotting symbol. See text for more details.}\label{fig9}

\end{center}
\end{figure}

The impact of the polarization on the knowledge of the galactic spectral index has been evaluated, and, for an experiments like TRIS, with a beam around $\sim$20$^{\circ}$ FWHM, we have shown that it is negligible everywhere, especially towards the galactic disk, even if there may be a suggestion that depolarization is less effective in the sky region $5^h\lesssim \alpha \lesssim 8^h$ (at $\delta=+42^{\circ}$), going from the anti-center towards the halo in the northern direction (see Figure \ref{fig5}).

TRIS-600 MHz data are accurate enough to allow a critical discussion of the zero level of the surveys at 150, 408, 820 and 1420 MHz, at least in the Northern Sky where our observations have been done. The results, obtained as described in section \ref{par_zero_maps} and discussed at the end of the previous section, are summarized in Table \ref{tab9}. The effect of this new assessment of the zero levels on the evaluation of the spectral index between 408 and 1420 MHz is also considered: a substantial steepening of the local spectral index, with $\Delta \beta \cong +0.2$, is observed towards the regions at high galactic latitude, at TRIS angular resolution.
This adjustment is important because these maps, noticeably the 408 MHz map \cite[]{haslam82}, are frequently used as a benchmark
in astrophysical and cosmological studies, e.g. for searching CMB anisotropies (e.g. Planck mission, \cite[]{tauber2004} and
SPOrt mission, \cite[]{cortiglioni04}) and in planning future observation of the diffuse radiation from space or from ground
stations.

%% The equation environment wil produce a numbered display equation.

%% The \notetoeditor{TEXT} command allows the author to communicate
%% information to the copy editor.  This information will appear as a
%% footnote on the printed copy for the manuscript style file.  Nothing will
%% appear on the printed copy if the preprint or
%% preprint2 style files are used.

%% The eqnarray environment produces multi-line display math. The end of
%% each line is marked with a \\. Lines will be numbered unless the \\
%% is preceded by a \nonumber command.
%% Alignment points are marked by ampersands (&). There should be two
%% ampersands (&) per line.

%% Included in this acknowledgments section are examples of the
%% AASTeX hypertext markup commands. Use \url without the optional [HREF]
%% argument when you want to print the url directly in the text. Otherwise,
%% use either \url or \anchor, with the HREF as the first argument and the
%% text to be printed in the second.

\acknowledgments  {\textit{Acknowledgements}}

We are grateful to Ettore Carretti for providing us the low resolution polarized maps at 610 and 820 MHz. Some of the results in this paper have been derived using the HEALPix package \cite{gorski2005}.

\clearpage

%% the file, uncomment the \plotone, \plottwo, and
%% If you need a layout that cannot be achieved with \plotone or
%% \plottwo, you can invoke the graphicx package directly with the
%% \includegraphics command or use \plotfiddle. For more information,
%% please see the tutorial on "Using Electronic Art with AASTeX" in the
%% documentation section at the AASTeX Web site,
%% http://www.journals.uchicago.edu/AAS/AASTeX.
%%
%% The examples below also include sample markup for submission of
%% supplemental electronic materials. As always, be sure to check
%% the instructions to authors for the journal you are submitting to
%% for specific submissions guidelines as they vary from

%% This example uses \plotone to include an EPS file scaled to
%% 80% of its natural size with \epsscale. Its caption
%% has been written to indicate that additional figure parts will be
%% available in the electronic journal.

\clearpage

\begin{deluxetable}{cccccc}
\tabletypesize{} \tablecaption{T-T plots: Spectral index between TRIS data at 600 MHz and 820 MHz}\tablewidth{0pt} \tablehead{
\colhead{$\alpha_0$(hh mm ss)} &\colhead{$\beta^{TT}_{Gal}$} & \colhead{$\chi^2/dof$} & \colhead{Q} & \colhead{$\beta_{Gal}^{TT,robust}$}& \colhead{$\Delta
\alpha$(deg)}} \startdata

09 00 00  & 3.0$\pm$0.1 & 0.48 & 0.85 & 2.97$\pm$0.01 & 105\\
10 00 00  & 3.1$\pm$0.1 & 0.42 & 0.89 & 3.06$\pm$0.02& 105 \\
11 00 00  & 3.1$\pm$0.2 & 0.50 & 0.84 & 3.13$\pm$0.04 & 105\\
\hline
19 00 00  & 2.7$\pm$0.1 & 1.17 & 0.32 & 2.79$\pm$0.02 & 75\\
20 00 00  & 2.73$\pm$0.09 & 1.15 & 0.33 & 2.75$\pm$0.02& 105 \\
20 24 00  & 2.8$\pm$0.1 & 0.95 & 0.46 & 2.75$\pm$0.02& 105 \\
21 00 00  & 2.7$\pm$0.1 & 0.98 & 0.44 & 2.69$\pm$0.02 & 90\\

\enddata

%\tablecomments{This table shows, for every value of right ascension $\alpha_0$, the spectral index $\beta_{TT}$, the parameters of the
%likelihood analysis ($\chi^2/dof$ and Q), the index derived using a robust fit technique $\beta_{robust}$ and finally the width $\Delta \alpha$
%(in degrees) of the right ascension interval considered for the T-T plot.}
\label{tab1}

\end{deluxetable}

%% \end{document}

%%
%% End of file `table.tex'.

\begin{deluxetable}{ccccc}
\tabletypesize{} \tablecaption{The brightness temperature of the Galaxy and the local spectral index between 600 and 820 MHz after correction
for polarization. The local galactic spectral index evaluated from TRIS data not corrected for polarization (see sec.\ref{tab2}) is reported in
the last column.}
 \tablewidth{0pt}\tablehead{ \colhead{$\alpha$(hh mm ss)} & \colhead{$T_{gal, 600}+\Delta T^{pol}_{600}$(K)} &
\colhead{$T_{gal, 820}+\Delta T^{pol}_{820}$(K)} & \colhead{$\beta^{pol}_{600-820}$} & \colhead{$\beta_{600-820}$} }\startdata

      00 00 00&      11.38$\pm$    0.07&      4.61$\pm$    0.03&      2.92$\pm$    0.03&     2.93$\pm$    0.03 \\
      01 00 00&      11.48$\pm$    0.07&      4.77$\pm$    0.03&      2.85$\pm$    0.03&      2.87$\pm$    0.03\\
      02 00 00&      12.32$\pm$    0.08&      5.17$\pm$    0.04&      2.80$\pm$    0.03&      2.80$\pm$    0.03\\
      03 00 00&      13.08$\pm$    0.08&      5.54$\pm$    0.04&      2.79$\pm$    0.03&      2.79$\pm$    0.02\\
      04 00 00&      13.60$\pm$    0.07&      5.63$\pm$    0.04&      2.85$\pm$    0.03&      2.89$\pm$    0.02\\
      05 00 00&      13.47$\pm$    0.08&      5.44$\pm$    0.04&      2.93$\pm$    0.03&      3.00$\pm$    0.02\\
      06 00 00&      11.87$\pm$    0.08&      4.73$\pm$    0.04&      2.99$\pm$    0.04&      3.04$\pm$    0.03\\
      07 00 00&      9.85$\pm$    0.08&      3.81$\pm$    0.04&      3.07$\pm$    0.04&      3.10$\pm$    0.03\\
      08 00 00&      7.95$\pm$    0.07&      3.03$\pm$    0.03&      3.12$\pm$    0.05&      3.14$\pm$    0.04\\
      09 00 00&      6.37$\pm$    0.07&      2.40$\pm$    0.03&      3.10$\pm$    0.05&      3.11$\pm$    0.05\\
      10 00 00&      5.72$\pm$    0.07&      2.21$\pm$    0.03&      3.08$\pm$    0.06&      3.07$\pm$    0.06\\
      11 00 00&      5.85$\pm$    0.07&      2.21$\pm$    0.03&      3.16$\pm$    0.06&      3.15$\pm$    0.06\\
      12 00 00&      6.39$\pm$    0.07&      2.34$\pm$    0.03&      3.22$\pm$    0.06&      3.21$\pm$    0.05\\
      13 00 00&      6.51$\pm$    0.07&      2.48$\pm$    0.03&      3.12$\pm$    0.05&      3.12$\pm$    0.05\\
      14 00 00&      6.97$\pm$    0.07&      2.73$\pm$    0.03&      3.04$\pm$    0.05&      3.06$\pm$    0.05\\
      15 00 00&      7.63$\pm$    0.07&      3.01$\pm$    0.03&      3.01$\pm$    0.04&      3.03$\pm$    0.04\\
      16 00 00&      8.37$\pm$    0.07&      3.29$\pm$    0.03&      3.02$\pm$    0.04&      3.02$\pm$    0.04\\
      17 00 00&      9.27$\pm$    0.07&      3.79$\pm$    0.03&      2.86$\pm$    0.04&      2.88$\pm$    0.03\\
      18 00 00&      10.98$\pm$    0.07&      4.85$\pm$   0.03&      2.66$\pm$    0.03&      2.67$\pm$    0.03\\
      19 00 00&      15.44$\pm$    0.08&      6.54$\pm$   0.03&      2.77$\pm$    0.02&      2.78$\pm$    0.02\\
      20 00 00&      23.26$\pm$    0.08&      9.72$\pm$   0.03&      2.82$\pm$    0.01&      2.82$\pm$    0.01\\
      20 24 00&      24.44 $\pm$   0.07&      10.38$\pm$  0.03&      2.78 $\pm$   0.01&      2.78$\pm$    0.01\\                                    21 00 00&      21.87$\pm$    0.07&      9.36$\pm$    0.03&      2.75$\pm$    0.01&      2.75$\pm$    0.01\\
      22 00 00&      15.56$\pm$    0.07&      6.65$\pm$    0.03&      2.76$\pm$    0.02&      2.76$\pm$    0.02\\
      23 00 00&      12.44$\pm$    0.07&      5.05$\pm$    0.03&      2.90$\pm$    0.03&      2.90$\pm$    0.03\\

\enddata
%% Text for table notes should follow after the \enddata but before
%% the \end{deluxetable}. Make sure there is at least one \tablenotemark
%% in the table for each \tablenotetext.
%\tablecomments{In this table we show the galactic temperature and the local galactic spectral index sampled with right ascension spacing
%$\Delta\alpha = 1^h$.}
\label{tab2}

\end{deluxetable}

%% \end{document}

%%
%% End of file `table.tex'.

\begin{deluxetable}{cccccc}
\tabletypesize{} \tablecaption{T-T plots: Spectral index between TRIS data at 600 MHz and 820 MHz corrected for polarization.}\tablewidth{0pt}
\tablehead{ \colhead{$\alpha_0$(hh mm ss)} &\colhead{$\beta_{Gal}^{TT}$} & \colhead{$\chi^2/dof$} & \colhead{Q} & \colhead{$\beta^{TT,robust}_{Gal}$}&
\colhead{$\Delta \alpha$(deg)}} \startdata

09 00 00  & 2.91$\pm$0.08 & 0.75 & 0.63 & 2.88$\pm$0.01 & 105\\
10 00 00  & 3.0$\pm$0.1 & 0.65 & 0.72 & 2.98$\pm$0.02& 105 \\
11 00 00  & 3.0$\pm$0.2 & 0.81 & 0.58 & 3.04$\pm$0.04 & 105\\
\hline
19 00 00  & 2.9$\pm$0.1 & 0.33 & 0.88 & 2.87$\pm$0.02 & 75\\
20 00 00  & 2.7$\pm$0.1 & 1.12 & 0.34 & 2.76$\pm$0.02 & 105 \\
20 24 00  & 2.8$\pm$0.1 & 0.92 & 0.49 & 2.76$\pm$0.02 & 105 \\
21 00 00  & 2.7$\pm$0.1 & 1.00 & 0.42 & 2.73$\pm$0.02 & 90\\

\enddata

%\tablecomments{See comments reported in tab.\ref{tab1}.}
\label{tab3}

\end{deluxetable}

%% \end{document}

%%
%% End of file `table.tex'.

\begin{deluxetable}{cccccc}
\tabletypesize{} \tablecaption{T-T plots: Spectral index between 408 MHz and 600 MHz}\tablewidth{0pt} \tablehead{ \colhead{$\alpha_0$(hh mm ss)}
&\colhead{$\beta_{Gal}^{TT}$} & \colhead{$\chi^2/dof$} & \colhead{Q} & \colhead{$\beta_{Gal}^{TT,robust}$}& \colhead{$\Delta \alpha$(deg)}} \startdata

09 00 00  & 2.58$\pm$0.09 & 0.68 & 0.69 & 2.66$\pm$0.01 & 105\\
10 00 00  & 2.5$\pm$0.1 & 0.55 & 0.80 & 2.51$\pm$0.02& 105 \\
11 00 00  & 2.3$\pm$0.2 & 0.46 & 0.87 & 2.35$\pm$0.04 & 105\\
\hline
19 00 00  & 2.80$\pm$0.08 & 0.76 & 0.57 & 2.77$\pm$0.02 & 75\\
20 00 00  & 2.83$\pm$0.08 & 0.90 & 0.51 & 2.76$\pm$0.02& 105 \\
20 24 00  & 2.82$\pm$0.09 & 0.95 & 0.47 & 2.75$\pm$0.02& 105 \\
21 00 00  & 2.7$\pm$0.1 & 0.80 & 0.56 & 2.67$\pm$0.02 & 90\\

\enddata

%\tablecomments{See comments reported in tab.\ref{tab1}.}
\label{tab4}

\end{deluxetable}

%% \end{document}

%%
%% End of file `table.tex'.

\begin{deluxetable}{cccccc}
\tabletypesize{} \tablecaption{T-T plots: Spectral index between 408 MHz and 820 MHz}\tablewidth{0pt} \tablehead{ \colhead{$\alpha_0$(hh mm ss)}
&\colhead{$\beta_{Gal}^{TT}$} & \colhead{$\chi^2/dof$} & \colhead{Q} & \colhead{$\beta_{Gal}^{TT,robust}$}& \colhead{$\Delta \alpha$(deg)}} \startdata

09 00 00  & 2.74$\pm$0.08 & 0.38 & 0.92 & 2.75$\pm$0.01 & 105\\
10 00 00  & 2.7$\pm$0.1 & 0.41 & 0.90 & 2.72$\pm$0.01& 105 \\
11 00 00  & 2.7$\pm$0.2 & 0.43 & 0.88 & 2.65$\pm$0.03 & 105\\
\hline
19 00 00  & 2.79$\pm$0.09 & 0.75 & 0.57 & 2.78$\pm$0.02 & 75\\
20 00 00  & 2.81$\pm$0.09 & 0.68 & 0.69 & 2.76$\pm$0.01& 105 \\
20 24 00  & 2.8$\pm$0.1 & 0.65 & 0.72 & 2.75$\pm$0.02& 105 \\
21 00 00  & 2.8$\pm$0.1 & 0.70 & 0.64 & 2.69$\pm$0.02 & 90\\

\enddata

%\tablecomments{See comments reported in tab.\ref{tab1}.}
\label{tab5}

\end{deluxetable}

%% \end{document}

%%
%% End of file `table.tex'.

\begin{deluxetable}{cccccc}
\tabletypesize{} \tablecaption{T-T plots: Spectral index between 600 MHz and 1420 MHz}\tablewidth{0pt} \tablehead{ \colhead{$\alpha_0$(hh mm
ss)} &\colhead{$\beta_{Gal}^{TT}$} & \colhead{$\chi^2/dof$} & \colhead{Q} & \colhead{$\beta_{Gal}^{TT,robust}$}& \colhead{$\Delta \alpha$(deg)}} \startdata

09 00 00  & 2.84$\pm$0.07 & 1.56 & 0.14 & 2.85$\pm$0.01 & 105\\
10 00 00  & 2.8$\pm$0.1 & 1.68 & 0.11 & 2.81$\pm$0.03& 105 \\
11 00 00  & 3.1$\pm$0.3 & 1.67 & 0.11 & 2.94$\pm$0.07 & 105\\
\hline
19 00 00  & 2.57$\pm$0.07 & 0.31 & 0.89 & 2.60$\pm$0.01 & 75\\
20 00 00  & 2.54$\pm$0.08 & 0.72 & 0.63 & 2.59$\pm$0.02& 105 \\
20 24 00  & 2.55$\pm$0.08 & 0.70 & 0.68 & 2.60$\pm$0.01& 105 \\
21 00 00  & 2.6$\pm$0.1 & 0.71 & 0.64 & 2.64$\pm$0.02 & 90\\

\enddata

%\tablecomments{See comments reported in tab.\ref{tab1}.}
\label{tab6}

\end{deluxetable}

%% \end{document}

%%
%% End of file `table.tex'.

\begin{deluxetable}{cccccc}
\tabletypesize{} \tablecaption{T-T plots: Spectral index between 820 MHz and 1420 MHz}\tablewidth{0pt} \tablehead{ \colhead{$\alpha_0$(hh mm
ss)} &\colhead{$\beta_{Gal}^{TT}$} & \colhead{$\chi^2/dof$} & \colhead{Q} & \colhead{$\beta_{Gal}^{TT,robust}$}& \colhead{$\Delta \alpha$(deg)}} \startdata

09 00 00  & 2.8$\pm$0.1 & 0.93 & 0.48 & 2.84$\pm$0.02 & 105\\
10 00 00  & 2.7$\pm$0.2 & 1.69 & 0.11 & 2.71$\pm$0.04& 105 \\
11 00 00  & 3.3$\pm$0.6 & 1.69 & 0.18 & 2.92$\pm$0.12 & 105\\
\hline
19 00 00  & 2.50$\pm$0.10 & 1.10 & 0.36 & 2.46$\pm$0.02 & 75\\
20 00 00  & 2.5$\pm$0.1 & 1.08 & 0.37 & 2.50$\pm$0.02& 105 \\
20 24 00  & 2.4$\pm$0.1 & 1.18 & 0.31 & 2.52$\pm$0.02& 105 \\
21 00 00  & 2.6$\pm$0.2 & 1.30 & 0.25 & 2.60$\pm$0.03 & 90\\

\enddata

%\tablecomments{See comments reported in tab.\ref{tab1}.}
\label{tab7}

\end{deluxetable}

%% \end{document}

%%
%% End of file `table.tex'.

\begin{deluxetable}{ccccc}
\tablewidth{0pt}\tabletypesize{} \tablecaption{T-T plots: summary. The galactic spectral index towards the galactic halo and at the peak of the galactic emission at $\delta=+42^{\circ}$.} \tablehead{\colhead{$(\nu_1, \nu_2)$} & \colhead{$\alpha=09^h$} &
\colhead{$\alpha=10^h$} & \colhead{$\alpha=11^h$} &\colhead{$\alpha=20^h 24^m$}} \startdata

& $(l,b)$ & $(l,b)$ & $(l,b)$ & $(l,b)$ \\
& ($179^{\circ}$, $41^{\circ}$)  & ($178^{\circ}$, $52^{\circ}$)  & ($172^{\circ}$, $63^{\circ}$) & ($80^{\circ}$, $3^{\circ}$) \\

\hline

$150-408$ & $2.1\pm 0.3$  & $2.2\pm 0.3$ & $2.2\pm 0.3$ & $\ldots$ \\
$150-600$ & $2.2\pm 0.2$  & $2.3\pm 0.2$ & $2.4\pm 0.2$ & $\ldots$\\

$150-820$ & $2.4\pm 0.2$  & $2.5\pm 0.2$  & $2.5\pm 0.2$ & $\ldots$ \\
$408-600$ & $2.6\pm 0.1$ & $2.5\pm 0.1$  & $2.3\pm 0.2$ & $2.82\pm 0.09$\\

$408-820$ & $2.74\pm 0.08$ & $2.7\pm 0.1$  & $2.7\pm 0.2$ & $2.8\pm 0.1$ \\
$600-820$\tablenotemark{a} & $2.91\pm 0.08$ & $3.0\pm 0.1$ & $3.0\pm 0.2$ & $2.8\pm 0.1$ \\

$408-1420$ & $2.73\pm 0.07$ & $2.6\pm 0.1$ & $2.4\pm 0.3$ & $2.63\pm0.07$ \\

$600-1420$ & $2.84\pm 0.07$ & $2.8\pm 0.1$ & $3.1\pm 0.3$ & $2.55\pm 0.08$\\

$820-1420$ & $2.8\pm 0.1$ & $2.7\pm 0.2$ & $3.3\pm0.6$ & $2.4\pm 0.1$\\

\enddata

\tablenotetext{a}{In this case TRIS data corrected for polarization are used.} \tablecomments{Here we show the frequencies ($\nu_1$ and $\nu_2$, in MHz) used to build the T-T plots. Also the galactic coordinates ($l$,$b$) are reported.} \label{tab8}

\end{deluxetable}

%% \end{document}

%%
%% End of file `table.tex'.

\begin{deluxetable}{ccccccc}
\tabletypesize{} \tablecaption{Suggested assessment of the zero level of continuum surveys at ($\alpha_m=10^h00^m,\delta_m=+42^{\circ}$).}
 \tablewidth{0pt}\tablehead{ \colhead{$\nu(MHz)$} & \colhead{$T^{(\alpha_m,\delta_m)}_{Gal}(K)$} &
\colhead{$T_{uers}(K)$} & \colhead{$T_b^{cmb}(K)$} & \colhead{$T^{(\alpha_m,\delta_m)}_{eval}$(K)} &
\colhead{$T_{map}^{(\alpha_m,\delta_m)}$(K)} & \colhead{$\Delta T(K)$} }\startdata

TRIS 600 & $5.63\pm 0.07$ & $0.94\pm 0.03$ & $2.82\pm 0.13$ & $\ldots$ & $\ldots$ & $\ldots$ \\

\hline
150\tablenotemark{a} & $139\pm 39 $ & $39\pm 2$  & $2.82\pm 0.13$ & $180\pm 39$ & 123$\pm 40$ & $+58\pm 39$ \\
408\tablenotemark{b} & $15.0\pm 0.6$  & $2.65\pm 0.09$  & $2.82\pm 0.13$  & $20.5\pm 0.6$ & $16.6\pm 3 $ & $+3.9\pm 0.6$ \\
820\tablenotemark{c} & $2.24\pm 0.07$ & $0.41\pm 0.01$ & $2.82\pm 0.13$  & $5.47\pm 0.15$ & $5.75\pm 0.6 $ & $-0.27\pm 0.15$ \\
1420\tablenotemark{d} & $0.51\pm 0.04$ & $0.094\pm 0.004$  & $2.82\pm 0.13$  & $3.43\pm 0.14$ & $3.31\pm 0.5$ & $+0.12 \pm 0.14$  \\

\enddata
%% Text for table notes should follow after the \enddata but before
%% the \end{deluxetable}. Make sure there is at least one \tablenotemark
%% in the table for each \tablenotetext.
\tablenotetext{a}{\cite{landecker70}}
\tablenotetext{b}{\cite{haslam82}}\tablenotetext{c}{\cite{berkhuijsen72}}\tablenotetext{d}{\cite{reich1986}}

\label{tab9}

\end{deluxetable}

%% \end{document}

%%
%% End of file `table.tex'.

\end{document}